\documentclass[a4paper,fleqn,amssymb,usenatbib,useAMS,usedcolumns]{mnras}
\usepackage{times} 
\usepackage{latexsym,amssymb,amsmath,amsfonts}
\usepackage{rotating} 
\usepackage{float}
\usepackage{graphicx}
\usepackage{natbib}
\usepackage{longtable}
\usepackage{url}
\usepackage{dcolumn}
\usepackage{textcomp}
\usepackage{multicol}
\usepackage{bm}
\usepackage{pdflscape}
\usepackage[T1]{fontenc}
\usepackage{ae,aecompl}
\usepackage[caption=false]{subfig}
\begin{document}
\newcommand{\hi}{\mbox{H\,{\sc i}}}
\newcommand{\mgii}{\mbox{Mg\,{\sc ii}}}
\newcommand{\mgi}{\mbox{Mg\,{\sc i}}}
\newcommand{\feii}{\mbox{Fe\,{\sc ii}}}
\newcommand{\mnii}{\mbox{Mn\,{\sc ii}}}
\newcommand{\crii}{\mbox{Cr\,{\sc ii}}}
\newcommand{\tii}{\mbox{Ti\,{\sc ii}}}
\newcommand{\sii}{\mbox{Si\,{\sc ii}}}
\newcommand{\znii}{\mbox{Zn\,{\sc ii}}}
\newcommand{\caii}{\mbox{Ca\,{\sc ii}}}
\newcommand{\nai}{\mbox{Na\,{\sc i}}}
\def\h2{$\rm H_2$}
\def\Nh2{$N$(H${_2}$)}
\def\chin{$\chi^2_{\nu}$}
\def\chiu{$\chi_{\rm UV}$}
\def\sys{J0441$-$4313~}
\def\lya{\ensuremath{{\rm Ly}\alpha}}
\def\lymana{\ensuremath{{\rm Lyman}-\alpha}}
\def\kms{km\,s$^{-1}$}
\def\cms{cm$^{-2}$}
\def\cc{cm$^{-3}$}
\def\zabs{$z_{\rm abs}$}
\def\zem{$z_{\rm em}$}
\def\nhi{$N$($\hi$)}
\def\ln{log~$N$}
\def\nh{$n_{\rm H}$}
\def\ne{$n_{e}$}
\def\21{21-cm}
\def\18{18-cm}
\def\ts{T$_{\rm s}$}
\def\te{T$_{\rm ex}$}
\def\ll{$\lambda\lambda$}
\def\l{$\lambda$}
\def\fc{$C_{f}$}
\def\fcoh{$f_{\rm c}^{\rm OH}$}
\def\fchi{$f_{\rm c}^{\textrm{H\,{\sc i}}}$}
\def\c21{$C_{21}$}
\def\mjb{mJy~beam$^{-1}$}
\def\taudv{$\int\tau dv$}
\def\taup{$\tau_{\rm p}$}
\def\ha{H\,$\alpha$}
\def\hb{H\,$\beta$}
\def\oi{[O\,{\sc i}]}
\def\oii{[O\,{\sc ii}]}
\def\oiii{[O\,{\sc iii}]}
\def\nii{[N\,{\sc ii}]}
\def\sii{[S\,{\sc ii}]}
\def\taudvl{$\int\tau dv^{3\sigma}_{10}$}
\def\taudv{$\int\tau dv$}
\def\vshift{$v_{\rm shift}$}
\def\wmg{$W_{\mgii}$}
\def\wfe{$W_{\feii}$}
\def\dgi{$\Delta (r-i)$}
\def\ebv{$E(B-V)$}
\def\noh{$N$(OH)}
\def\ohi{[OH]$/$[H\,{\sc i}]}
%
%
\title[Cold gas towards red quasars]{
{uGMRT search for cold gas at $z\sim1-1.4$ towards red quasars
}
\author[R. Dutta et al.]{R. Dutta$^1$ \thanks{E-mail: rdutta@eso.org}, R. Srianand$^2$, N. Gupta$^2$, R. Joshi$^3$ \\ 
$^1$ European Southern Observatory, Karl-Schwarzschild-Str. 2, D-85748 Garching Near Munich, Germany \\
$^2$ Inter-University Centre for Astronomy and Astrophysics, Post Bag 4, Ganeshkhind, Pune 411007, India \\
$^3$ Kavli Institute for Astronomy and Astrophysics, Peking University, Beijing 100871, People's Republic of China} 
}
\date{Accepted. Received; in original form }
\pubyear{}
\maketitle
\label{firstpage}
\pagerange{\pageref{firstpage}--\pageref{lastpage}}
%
%
\begin {abstract}  
\par\noindent
We present results from our search for \hi\ \21\ and OH \18\ absorption at $z\sim1-1.4$ towards red quasars showing strong \mgii\ absorption using uGMRT.
The quasars J1501$+$1822 and J1521$+$5508 show multiple strong associated \mgii\ absorption at $z\sim1.1$ and signature of reddening in their optical
spectra. We report the detection of \hi\ \21\ absorption towards J1521$+$5508 at the systemic redshift of the quasar, with \nhi\ $=(1.2\pm0.2)\times10^{20}$\,\cms\ 
for spin temperature of 100 K and unit covering factor. The \hi\ \21\ absorption is offset from the blueshifted strong \mgii\ absorbers by $\gtrsim$1500\,\kms. 
We do not detect \hi\ \21\ absorption at the redshift of the associated \mgii\ absorption and at the systemic redshift towards J1501$+$1822. We argue that 
lack of one-to-one correspondence between \mgii\ and \hi\ \21\ absorption could be related with clumpiness of the neutral gas and the radio and optical 
sightlines probing different volume of the gas. We find that the presence of strong associated \mgii\ absorption and reddening in the optical spectra of 
the quasars lead to an increased detection rate of associated \hi\ \21\ absorption at $z\ge$1. We also report non-detection of intervening OH absorption 
(\ohi\ $\le$(1-4)$\times10^{-8}$) at $z$ = 1.3 towards two red quasars, J0850$+$5159 and J0852$+$3435, which show strong metal and \hi\ \21\ absorption and 
the 2175\,\AA\ dust extinction bump. 
\end {abstract}  
%
%
%
%
\section{Introduction} 
\label{sec_introduction}  
\hi\ \21\ absorption towards radio-loud quasars can be used to probe the cold neutral gas in intervening galaxies, 
usually selected via the presence of \lymana\ or \mgii\ absorption in the quasar's optical spectra
\citep[e.g.][]{briggs1983,lane2000,curran2010,kanekar2009,kanekar2014,gupta2009,gupta2012,srianand2012,dutta2017a,dutta2017b,dutta2017c}.
The average detection rate of \hi\ \21\ absorption in strong \mgii\ systems is $\sim$20\% over $0.5<z<1.5$, and has 
been found to increase with equivalent width of \mgii\ and \feii\ absorption \citep{gupta2012,dutta2017b}, while the 
detection rate in $2<z<3$ damped \lymana\ absorbers is $\sim$10-20\% \citep{srianand2012,kanekar2014}. The neutral gas 
in the environment of radio-loud active galactic nuclei (AGNs) can be probed by \hi\ \21\ absorption as well
\citep[e.g.][]{vangorkom1989,gupta2006,chandola2011,curran2013a,allison2014,gereb2015,maccagni2017,aditya2018a,aditya2018b,dutta2018,dutta2019,grasha2019}.
Such studies find a typical detection rate of $\sim$30\% in $z<1$ radio galaxies. However, searches for \hi\ \21\ 
absorption associated to $z>1$ AGNs have been mostly unsuccessful ($\le$10\% detection rate). This could reflect 
either a lack of neutral gas along the line of sight due to geometric effects \citep[e.g.][]{gupta2006} or stronger 
ionization by the high-$z$ AGNs compared to the low-$z$ samples \citep[e.g.][]{curran2008,grasha2019}. 

Complementary to \hi\ \21\ absorption, OH \18\ absorption lines can be used to study the cold molecular phase of the gas. 
The OH radical is one of the common constituents of diffuse and dense molecular clouds and one of the best tracers of molecular 
hydrogen in the Galaxy \citep{liszt1996,liszt1999}. In addition, the \hi\ \21\ and OH \18\ absorption lines, if detected at
high significance level, can be used to place tight constraints on the variations of fundamental constants of physics \citep{uzan2003}.
However, only six OH absorbers at cosmologically significant redshifts in radio wavelengths are known to date \citep{chengalur1999,chengalur2003,kanekar2002,kanekar2005,gupta2018}. 

Quasars obscured by dust, either in intervening galaxies along the line-of-sight or in the quasar host galaxy itself, could present 
optimal sightlines to detect cold gas associated with the dust via \hi\ and metal absorption. Such quasars could be identified on the 
basis of their red optical, infrared or optical-to-infrared colours and in radio-selected samples \citep{gregg2002,ross2015,glikman2018}. 
We note, however, that the red colours of quasars could also arise due to their intrinsically red spectral energy distribution (SED), 
for example due to synchrotron emission from the radio jets \citep{srianand1997,whiting2001}. 

There have been observations of \hi\ \21\ absorption arising from dusty intervening galaxies that cause reddening in the optical 
spectra of the background quasars \citep{srianand2008,dutta2017b}. In case of associated \hi\ \21\ absorption, \citet{carilli1998} 
have found a high ($\sim$80\%) detection rate in a sample of five $z\sim0.7$ radio-selected optically-faint quasars \citep[see also][]{ishwarachandra2003}. 
\citet{chandola2017} have found that $z\le0.2$ radio galaxies with redder mid-infrared colours from Wide-field Infrared Survey Explorer 
\citep[WISE;][]{wright2010} show a higher detection rate of \hi\ \21\ absorption. Further, \citet{curran2006} have found that the 
molecular fraction and line strength of associated OH absorbers are related with the optical-to-infrared colour of the quasars. 

However, there have also been studies that find low detection rates of associated \hi\ \21\ absorption in obscured quasars. 
\citet{yan2016} have found a $\sim$20\% detection rate in obscured AGNs at $0.2<z<1.5$, selected on basis of optical-to-near-infrared 
colours, although their results could be limited by presence of radio frequency interference (RFI) and absence of accurate 
spectroscopic redshifts. Recently, \citet{glowacki2019} have found a detection rate of $\sim$12.5\% ($\sim$25\% for sources 
with spectroscopic redshifts) in a sample of $0.4<z<1$ quasars selected on the basis of their radio flux and either faint 
optical magnitude or red WISE mid-infrared colour. All the three \hi\ \21\ absorption detected by them are from the sub-sample 
of optically-faint quasars. However, the lack of \hi\ \21\ absorption in their WISE mid-infrared colour-selected sub-sample is
not statistically significant.

Here we explore an alternative method to select dust-obscured red quasars to search for \hi\ \21\ and OH \18\ absorption
$-$ quasars showing optical reddening and strong \mgii\ absorption. One expects dust to be associated with metal-enriched 
gas, and \ebv\ or reddening of quasars has been found to depend on the equivalent width of strong \mgii\ absorption \citep{budzynski2011}. 
\citet{dutta2017b} have demonstrated that intervening \hi\ \21\ absorption arises on an average in systems with stronger metal
absorption and towards quasars which show higher reddening [i.e. \ebv] due to dust associated with the \hi\ and metal absorption
in their optical spectra. We investigate here whether red quasars showing strong associated \mgii\ absorption give rise to \hi\ 
\21\ absorption as well. According to  hierarchical models of gas-rich major merger-driven galaxy formation, in the initial black 
hole growth phase fueled by gas inflow, the nuclear region is buried in dust \citep{sanders1996,hopkins2008}. Associated \hi\ \21\ 
absorption from red quasars is thus an useful tool to probe feedback processes in this dust-enshrouded phase of quasar evolution. 
Recently, we have obtained high detection rate ($\sim$84\%) of \hi\ \21\ absorption in $z\le$0.2 radio-loud galaxy mergers 
\citep{dutta2018,dutta2019}. Dust-reddened quasars can allow us to probe the merger phase at higher redshifts as well.

To address the paucity of \hi\ \21\ and OH \18\ absorption at high-$z$, we have started to search for these lines in previously 
unexplored redshift ranges towards $z>1$ red quasars, using the newly commissioned upgraded Giant Metrewave Radio Telescope 
\citep[uGMRT;][]{gupta2014} receivers. These receivers offer almost continuous frequency coverage over $120-1450$\,MHz, thus 
allowing one to search for absorption lines in redshift ranges that were not previously accessible to radio interferometers in
relatively RFI-free environment, e.g. \hi\ \21\ at $z\sim0.7-1.2$ and OH \18\ at $z\sim1-1.6$. Such searches would be
useful to plan future observations with the Square Kilometre Array precursors, ASKAP \citep{schinckel2012} and MeerKAT \citep{booth2012}. 
Note that although the Green Bank Telescope currently offers broad frequency coverage below 1\,GHz, it is a single dish and 
hence more affected by RFI. Here we present the results from our search for intervening OH \18\ main absorption lines at $z=1.3$ 
towards the red quasars $-$ J085042.24$+$515911.6 (hereafter J0850$+$5159) and J085244.74$+$343540.4 (hereafter J0852$+$3435); 
and our search for associated \hi\ \21\ absorption at $z=1.1$ towards the red quasars $-$ J150129.87$+$182221.1 (hereafter 
J1501$+$1822) and J152134.17$+$550857.2 (hereafter J1521$+$5508). The details of the sources and observations are presented 
in Section~\ref{sec_obs}. The results on individual sources are presented in Section~\ref{sec_results}. These are discussed 
and summarized in Section~\ref{sec_discussion}.
%
%
\section{Sample \& Observations}
\label{sec_obs}
\subsection{Source Selection}
\label{sec_sample}
\subsubsection{Sources for OH \18\ absorption search}
\label{sec_sample1}
We selected J0850$+$5159 and J0852$+$3435 to search for OH \18\ absorption because they both show strong \hi\ \21\ and metal absorption 
lines, dust reddening [\ebv\ $\sim$0.3-0.4] and the 2175\,\AA\ ultraviolet (UV) extinction bump \citep{srianand2008}. Note that the 
broad absorption bump at 2175\,\AA\ is a spectroscopic feature due to extinction by interstellar dust, most likely polycyclic aromatic 
hydrocarbons, that is seen in the Milky Way, Magellanic Clouds, as well as high-$z$ absorption-line systems \citep{draine2003,wang2004,jiang2011}.
Hence, while the reason for the reddening of the optical continuum of the quasar could be degenerate (see Section~\ref{sec_introduction}), 
the 2175\,\AA\ bump confirms the presence of dust unambiguously.

The above two quasars were part of a systematic survey of \hi\ \21\ absorption in $z\sim1.3$ strong \mgii\ systems \citep{gupta2009}. Strong 
absorption from \mgii\ [rest equivalent width (REW) of \mgii\,\l2796, \wmg\ = 4.6\,\AA\ and 2.9\,\AA\ for J0850$+$5159 and J0852$+$3435, 
respectively], \feii\ (REW of \feii\,\l2600, \wfe\ = 2.3\,\AA\ and 2.1\,\AA\ for J0850$+$5159 and J0852$+$3435, respectively), as well as 
weaker transitions of \mnii, \znii, \crii\ and \tii, are detected at $z=1.3$ in the Sloan Digital Sky Survey \citep[SDSS;][]{york2000} 
spectra of the quasars. In addition, the 9.7$\micron$ silicate feature is detected in absorption from the $z=1.3$ system towards J0852$+$3435 
\citep{kulkarni2011}. The background radio sources are relatively weak, with 1.4~GHz flux density of 63\,mJy and 69\,mJy for J0850$+$5159 
and J0852$+$3435, respectively.

In the \hi\ \21\ absorption survey of $z\sim1.3$ strong \mgii\ systems, there is another system associated with a 2175\,\AA\ bump 
towards the red quasar J095631.05$+$404628.2 \citep[J0956$+$4046;][]{gupta2012}. This system is very similar to J0850$+$5159 and J0852$+$3435 
in terms of metal line strength and reddening properties. However, the background radio source in this case is resolved into two components 
in the GMRT 610~MHz image and no \hi\ \21\ absorption is detected towards it, leading to constraint on the spin temperature (\ts) and covering 
factor (\fchi) of the gas, \ts$/$\fchi\ $>$7800\,K \citep[for details see section 5.2 of][]{gupta2012}. Hence, we did not consider this 
system for OH \18\ search.

\subsubsection{Sources for \hi\ \21\ absorption search}
\label{sec_sample2}
We identified the red quasars, J1501$+$1822 and J1521$+$5508, from the SDSS Data Release 12 \citep[DR12;][]{alam2015}, specifically 
using the \mgii\ catalog of \citet{zhu2013}. We searched for strong \mgii\ absorption [\wmg\ $\ge1$\,\AA] whose redshifted \hi\ \21\ 
line can be observed with uGMRT, towards radio-loud quasars [1.4\,GHz flux $\ge100$\,mJy in the Faint Images of the Radio Sky at 
Twenty-Centimeters \citep[FIRST;][]{white1997} catalog]. We restricted to sources where the optical and radio sightlines match within 
$1''$ radius. We then selected those quasars which have $r-i$ colour excess [$\Delta (r-i)$] greater than $3\sigma$ of the distribution, 
where $\Delta (r-i)$ is calculated by comparing the $r-i$ colour of the quasar to the median $r-i$ colour for SDSS quasars at the same 
redshift. This led to a sample of fifteen red quasars. We visually inspected the optical spectra of these quasars, and after removing 
systems with false identification of \mgii\ or incorrect estimate of \wmg, Broad Absorption Line (BAL) quasars, and double radio 
sources (to avoid ambiguity between optical and radio sightlines), we were left with eight quasars. Five of them show intervening \mgii\ 
absorption over $z\sim$0.6-2.0 and three show associated (within $\sim$3000\,\kms\ of the systemic redshift) \mgii\ absorption at $z\sim$1.1.

We fitted the quasar SDSS spectra by using the composite spectra of \citet{selsing2016} and \citet{vandenberk2001}, reddened by 
the Milky Way, Small Magellanic Cloud (SMC), Large Magellanic Cloud (LMC) or LMC2 supershell extinction curves \citep{gordon2003}. 
Both the composite spectra give similar results. We assume that the reddening is caused due to dust present in the strong \mgii\
absorption. For details of the SED fitting procedure we refer to \citet{srianand2008}. The reddening of the quasars thus obtained 
is in the range of \ebv\ $\sim$0.1-0.3, which is significant at $\ge2\sigma$ on comparing to control sample of quasars at 
the same redshift (see Section~\ref{sec_j0852} for further details). Of the above eight quasars, six have been previously searched 
for \hi\ \21\ absorption at the redshift of the strong \mgii\ system, with two detections \citep{ishwarachandra2003,gupta2009,kanekar2009,dutta2017b}
\footnote{One of the quasars, J092136.23$+$621552.1, showing intervening \mgii\ and \hi\ \21\ absorption \citep{dutta2017b}, 
has been searched for and not detected in associated \hi\ \21\ absorption \citep{aditya2016}. However, there is no associated 
\mgii\ absorption (\wmg\ $\le$1\,\AA) detected in the SDSS spectrum as well.}. Results of the \hi\ \21\ absorption search 
towards the remaining two quasars, showing multiple strong associated \mgii\ absorption at $z\sim1.1$, are presented in this work. 
\subsection{Observations}
\label{sec_observations}  
The radio spectral line searches were conducted using Band 4 (550$-$850 MHz) receivers of the uGMRT and the GMRT software back-end 
as the correlator (Proposal IDs: 32\_050, 33\_100, 34\_022, 34\_034). The details of the observations are listed in Table~\ref{tab:obs}. 
The observations were carried out using 2, 4 or 16 MHz baseband bandwidth split into 512 channels. During the observations, standard 
calibrators were regularly observed for flux density, bandpass and gain calibration. The data were acquired in parallel hand correlations,
and were reduced using the National Radio Astronomy Observatory (NRAO) Astronomical Image Processing System ({\sc aips}) following standard 
procedures \citep[see][for details]{dutta2016}.
\begin{table*}
\caption{uGMRT observation log.}
\centering
\begin{tabular}{cccccccc}
\hline
Quasar & \zem\ & \zabs\ & Date & Time & Central   & Spectral & Channel \\
       &       &        &      &      & Frequency & coverage & width   \\
       &       &        &      & (h)  & (MHz)     & (\kms)   & (\kms)  \\
(1)    & (2)   & (3)    & (4)  & (5)  & (6)       & (7)      & (8)     \\
\hline
J0850$+$5159 & 1.894 & 1.3265                 & 2017 July 02, 03 & 8.4  & 716.3 & 1745 & 3.4  \\
J0852$+$3435 & 1.652 & 1.3095                 & 2017 July 22, 23 & 25.7 & 721.5 & 1730 & 3.4  \\
             &       &                        & 2018 August 11, 13, 19  &    &     &     &    \\
J1501$+$1822 & 1.186 & 1.1859, 1.1818         & 2018 January 19  & 5.8  & 650.4 & 1920 & 3.8  \\
J1521$+$5508 & 1.070 & 1.0450, 1.0593, 1.0654 & 2018 January 01  & 6.1  & 692.2 & 7200 & 14.1 \\
             &       &                        & 2018 August 12   & 5.3  & 686.1 & 910  & 1.8  \\
\hline
\end{tabular}
\label{tab:obs}
\begin{flushleft}
{\it Notes.} Column 1: quasar name. Column 2: redshift of quasar. Column 3: redshift of strong \mgii\ absorption.
Column 4: observation date. Column 5: total time on-source in hours. Column 6: central observing frequency in MHz. 
Column 7: spectral coverage in \kms. Column 8: channel width in \kms. \\
\end{flushleft}
\end{table*}
%
%
%
\section{Results}
\label{sec_results}
\begin{figure*}
\subfloat[J0850$+$5159]{\includegraphics[width=0.35\textwidth, angle=90]{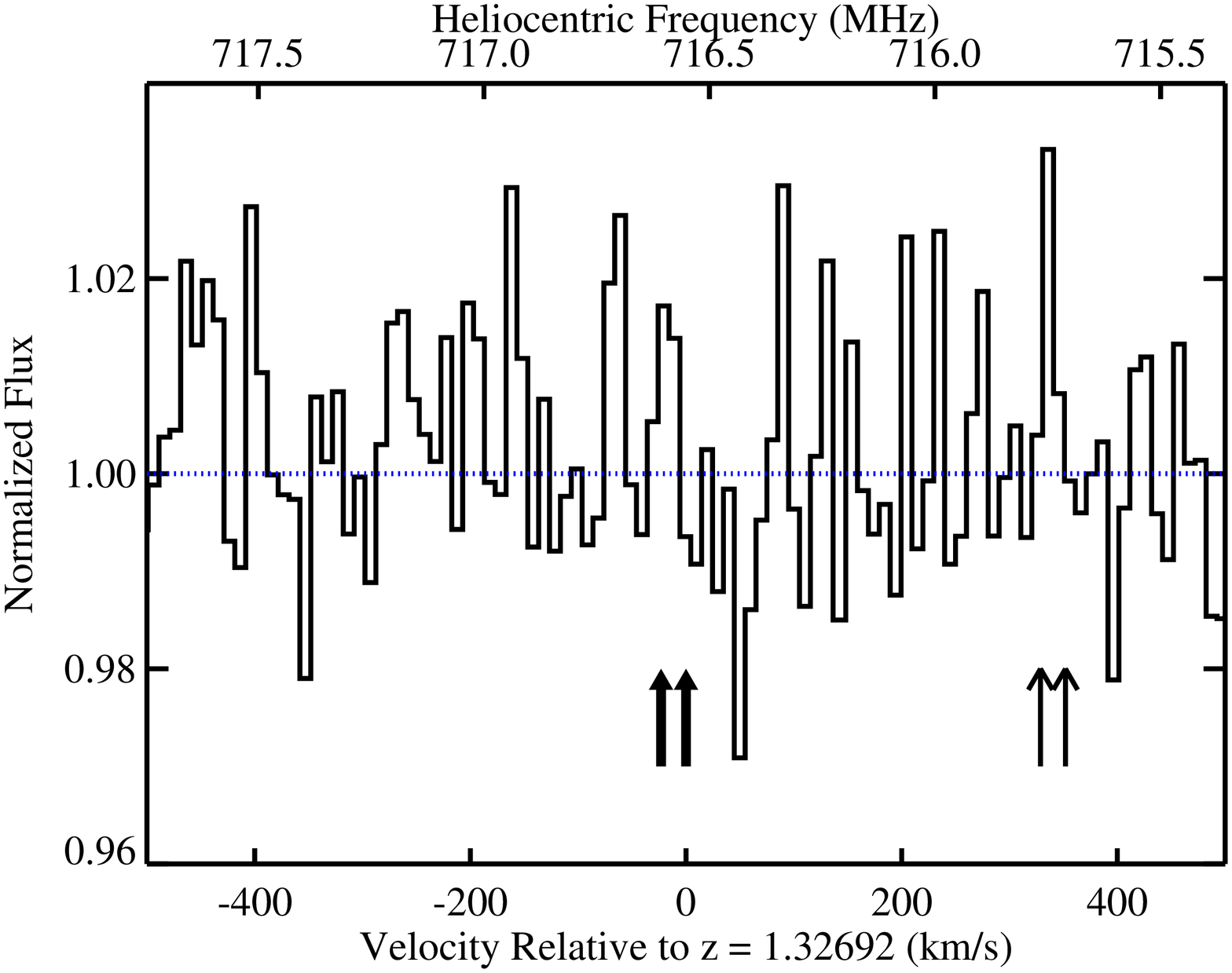} } 
\subfloat[J0852$+$3435]{\includegraphics[width=0.35\textwidth, angle=90]{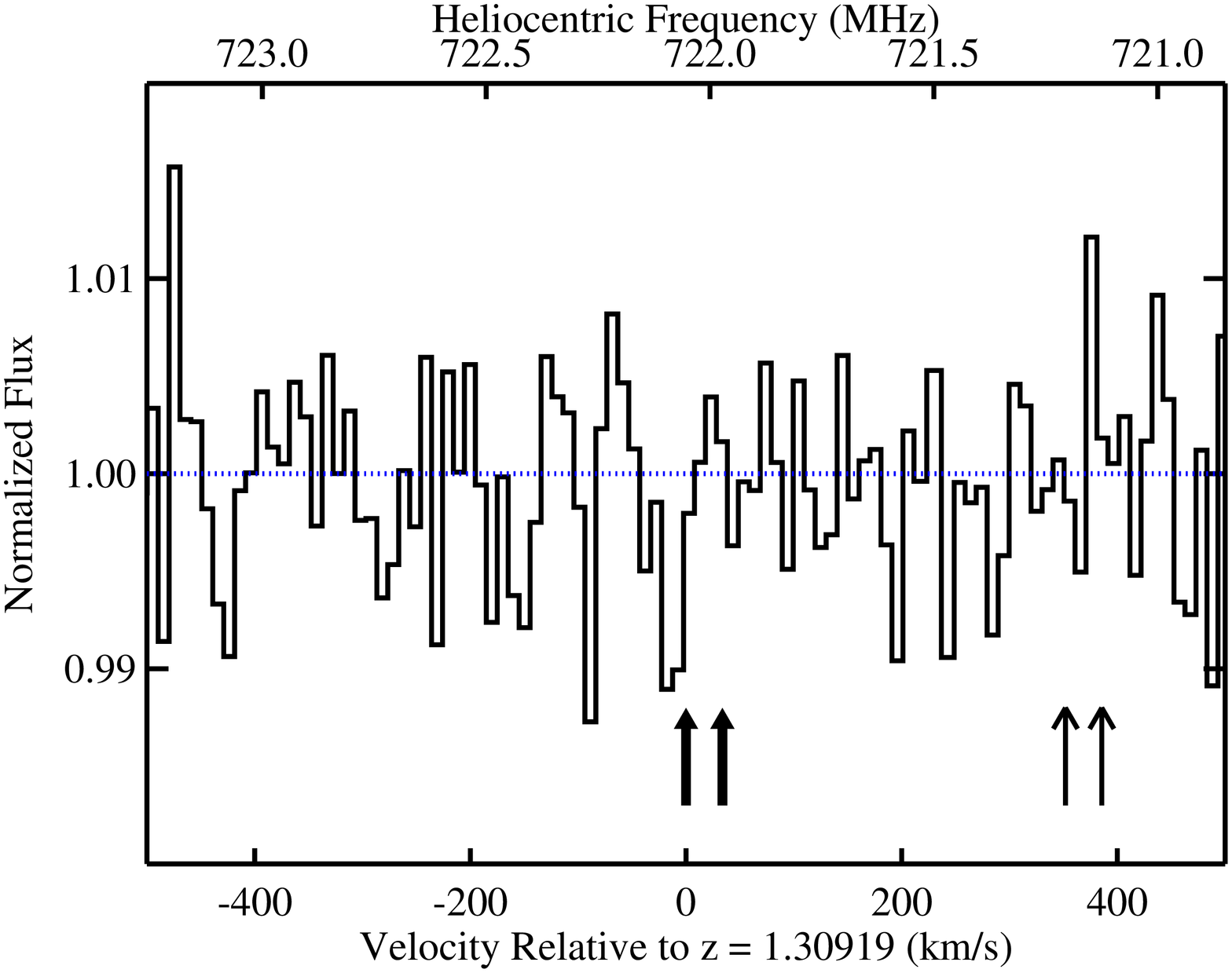} } \vspace{0.1cm}
\subfloat[J1501$+$1822]{\includegraphics[width=0.35\textwidth, angle=90]{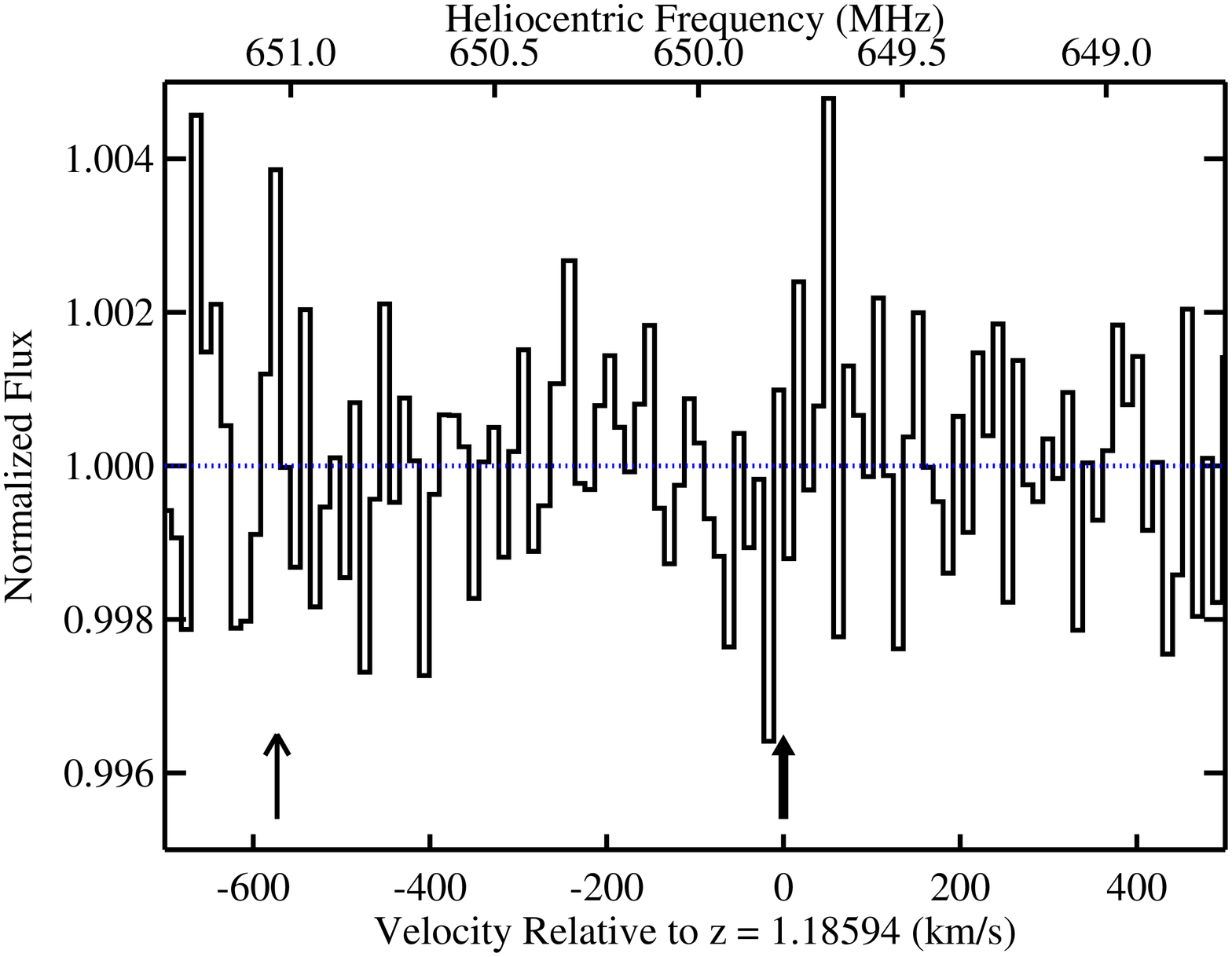} } 
\subfloat[J1521$+$5508]{\includegraphics[width=0.35\textwidth, angle=90]{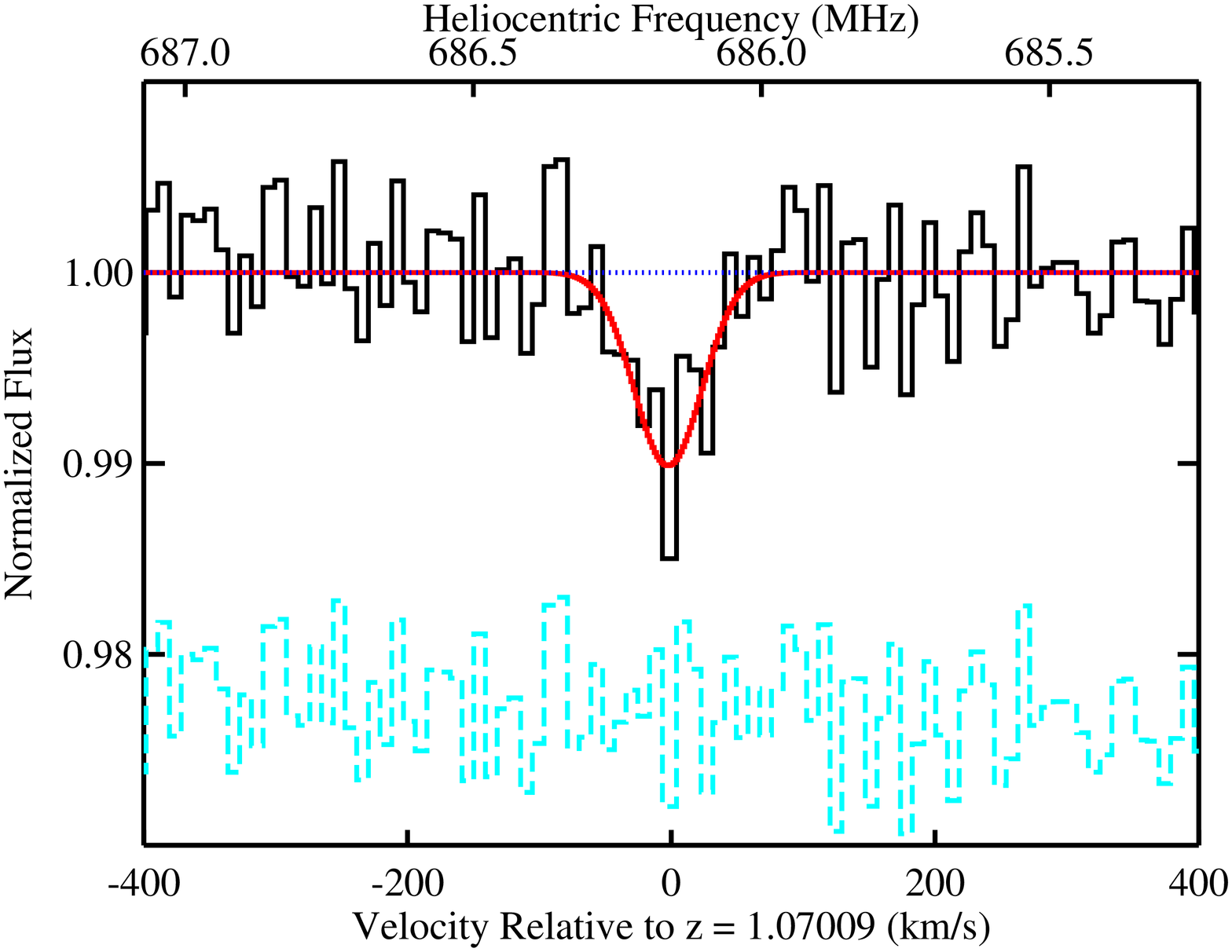} } 
\caption{Absorption spectra towards the red quasars studied here. The spectra have been smoothed to 10\,\kms\ for display purpose.
(a) OH \18\ absorption spectrum towards J0850$+$5159. The thick (thin) arrows indicate the expected positions of the redshifted 1667 MHz (1665 MHz) 
line from the \hi\ \21\ absorption components at $z = 1.32692$ and $z = 1.32674$. 
(b) OH \18\ absorption spectrum towards J0852$+$3425. The thick (thin) arrows indicate the expected positions of the redshifted 1667 MHz (1665 MHz) 
line from the \hi\ \21\ absorption components at $z = 1.30919$ and $z = 1.30945$.
(c) \hi\ \21\ absorption spectrum towards J1501$+$1822. The thick (thin) arrow indicates the expected position of the redshifted \hi\ \21\ line from 
the \mgii\ absorption at $z = 1.18594$ ($z = 1.18177$).
(d) \hi\ \21\ absorption detected towards J1521$+$5508. The best-fitting Gaussian is overplotted in red solid line. The residual from the fit 
is shown in blue dashed line, shifted arbitrarily on the $y$-axis. The velocity scale is with respect to the systemic redshift of the quasar. 
As discussed in the text, no \hi\ \21\ absorption is detected at the redshifts of the \mgii\ absorption.
}
\label{fig:absspectra}
\end{figure*}

The results derived from the uGMRT observations are listed in Table~\ref{tab:results}. All the radio sources are compact in the 
continuum maps (spatial resolution of $\sim4-5''$). The absorption spectra (see Fig.~\ref{fig:absspectra}) are extracted at the 
location of the peak continuum flux density of the radio sources. Below we discuss the results for each source in detail.
\begin{table}
\caption{Results from the radio spectral line observations.}
\centering
\begin{tabular}{p{1.4cm}p{0.8cm}p{0.8cm}p{1.5cm}p{2.2cm}}
\hline
Quasar & Peak         & Spectral     & \centering $\tau_{\rm p}$ & \taudv \\
       & Flux         & rms          &                           &        \\
       & Density      &              &                           &        \\
       & (mJy         & (mJy         &                           & (\kms) \\
       & beam$^{-1}$) & beam$^{-1}$) &                           &        \\
(1)    & (2)          & (3)          & \centering (4)            & (5)    \\
\hline
J0850$+$5159 & 65  & 1.1 & $\le$0.017      & $\le$0.224 (OH)      \\
J0852$+$3435 & 58  & 0.4 & $\le$0.007      & $\le$0.092 (OH)      \\
J1501$+$1822 & 243 & 0.7 & $\le$0.003      & $\le$0.144 (\hi)    \\
J1521$+$5508 & 195 & 1.0 & 0.021$\pm$0.005 & 0.65$\pm$0.09 (\hi) \\
\hline
\end{tabular}
\label{tab:results}
\begin{flushleft} 
{\it Notes.}
Column 1: quasar name. 
Column 2: peak flux density in \mjb. 
Column 3: spectral rms in \mjb\ per $\sim$3\,\kms. 
Column 4: peak optical depth in case of detection or 1$\sigma$ upper limit in case of non-detections.
Column 5: integrated optical depth in case of detection or 3$\sigma$ upper limit in case of non-detections.
Note that the \taudv\ upper limits corresponding to the OH 1667 MHz line towards J0850$+$5159 and J0852$+$3435
have been estimated for velocity width of 5\,\kms, while the \taudv\ upper limit for the \hi\ \21\ line
towards J1501$+$1822 have been estimated for velocity width of 100\,\kms.
\end{flushleft}
\end{table}
\begin{table}  
\centering
\caption{Parameters obtained from SED fit to the quasars.}
\begin{tabular}{p{1.4cm}p{1.4cm}p{1.5cm}p{1.2cm}p{1.2cm}}
\hline
Quasar & Dust  & \ebv\            & \nhi\                    & ($\kappa$ \ts)   \\
       & Model &                  & $\times1/\kappa~10^{21}$ & $\times1/$\fchi\ \\
       &       &                  & (\cms)                   & (K)              \\
 (1)   & (2)   & \centering (3)   & (4)                      & (5)              \\
\hline
J0850$+$5159 & LMC2    & 0.27 $\pm$ 0.01 & 5.2 $\pm$ 1.0 & 191$^{+139}_{-73}$     \\
J0852$+$3435 & LMC2    & 0.36 $\pm$ 0.01 & 7.0 $\pm$ 1.3 & 534$^{+257}_{-171}$    \\
J1501$+$1822 & SMC Bar & 0.24 $\pm$ 0.01 & 8.7 $\pm$ 1.0 & $\ge3\times10^4$       \\
J1521$+$5508 & SMC Bar & 0.15 $\pm$ 0.01 & 5.3 $\pm$ 0.8 & 4473$^{+1503}_{-1137}$ \\
\hline
\end{tabular}
\begin{flushleft} {\it Notes.}
Column 1: quasar name. Column 2: Best-fitting dust extinction model. Column 3: Best-fitting \ebv.  
Column 4: \nhi\ in units of $1/\kappa~10^{21}$ (\cms) obtained from the relation between \nhi\ and $A_V$ \citep{gordon2003}.
$\kappa$ is the dust-to-gas ratio relative to the best-fitting dust model.
Column 5: ($\kappa$ \ts)$/$\fchi\ (K) obtained by comparing the \nhi\ from SED fit with the observed \hi\ \21\ \taudv.  
The SED fit parameters for J0850$+$5159 and J0852$+$3435 are taken from \citet{srianand2008}.
\end{flushleft}
\label{tab:sed}
\end{table}
\subsection{OH \18\ absorption towards J0850$+$5159}
\label{sec_j0850}
\citet{srianand2008} have modeled the SED of the quasar J0850$+$5159 using the SDSS composite spectrum of \citet{vandenberk2001} 
and different dust extinction curves redshifted to the rest-frame of the intervening \mgii\ absorption. The LMC2 supershell 
extinction curve is found to give best fit to the quasar SED, with \ebv\ $\sim$0.3. The SED modeling and the $A_V$ versus \nhi\ 
relationship in the LMC \citep{gordon2003} gives \nhi\ $\sim5\times10^{21}\kappa^{-1}$\,\cms, where $\kappa$ is the dust-to-gas 
ratio relative to the LMC. The results from SED fitting are listed in Table~\ref{tab:sed}. The \hi\ \21\ absorption consists of 
two components (full-width-at-half-maximum, FWHM = 24\,\kms\ and 49\,\kms) separated by $\sim$23\,\kms, with an integrated optical 
depth of 15\,\kms.  

We do not detect absorption from the two OH \18\ main lines in this system (see panel (a) of Fig.~\ref{fig:absspectra}). We reach
a 3$\sigma$ optical depth limit for the 1667 MHz line of 0.224\,\kms\ over a velocity width of 5\,\kms. This translates to a column
density upper limit of \noh\ $\le2\times10^{14}$ (\te/3.5~K) (1/\fcoh) \cms, where \te\ is the excitation temperature and \fcoh\ is 
the covering factor \citep[following][]{liszt1996,stanimirovic2003}. \te\ = 3.5 K is the peak of the lognormal function fitted to the 
\te\ distribution of Galactic OH absorbers \citep{li2018}. The radio source is compact in the Very Long Baseline Array (VLBA) 20-cm 
sub-arcsecond-scale map presented in \citet{gupta2012}, which recovers all the arcsecond-scale flux. Hence, we take \fcoh\ to be unity. 
Using the \nhi\ obtained from SED fitting discussed above and assuming $\kappa$ = 1, we can put a limit on the abundance ratio 
\ohi\ $\le4\times10^{-8}$.
\subsection{OH \18\ absorption towards J0852$+$3435}
\label{sec_j0852}
The SED of the quasar J0852$+$3435, similar to J0850$+$5159, is best fit with LMC2 extinction curve and \ebv\ $\sim$0.4 \citep{srianand2008}. 
The \nhi\ derived from SED modeling is $\sim7\times10^{21}\kappa^{-1}$\,\cms. The \hi\ \21\ absorption from the \mgii\ system comprises a 
narrow (FWHM = 23\,\kms) and a broad (FWHM = 63\,\kms) component, separated by $\sim$34\,\kms. The total optical depth of the absorption is 7\,\kms.  

We initially detected a tentative absorption feature near the expected position of the redshifted OH \18\ main lines from this system in 2017. 
Hence, we obtained deeper observations of this system in 2018 (see Table~\ref{tab:obs}). We observed this system for a total of $\sim$26\,h 
on-source, reaching a spectral rms of 0.4\,\mjb\ per 3\,\kms\ channel. However, we do not detect any significant absorption feature in the 
final spectrum. The spectrum shown in panel (b) of Fig.~\ref{fig:absspectra} is obtained by coadding the noise-weighted normalized spectra 
from the five different observing runs. The tentative absorption feature is not produced consistently in all the runs as well as in the two 
different polarizations. The initial tentative feature could have been due to narrow sporadic low-level RFI (at 721.5\,MHz), that is not
significant in the final coadded spectra, due to the Doppler shift caused by heliocentric motion of the Earth between the different observing runs.

The final spectrum has a 3$\sigma$ optical depth sensitivity for the 1667 MHz line of 0.092\,\kms\ over a velocity width of 5\,\kms. This 
corresponds to \noh\ $\le1\times10^{14}$ (\te/3.5~K) (0.68/\fcoh) \cms. We estimate \fcoh\ = 0.68 based on the ratio of the flux density of 
the strongest component of the radio source detected in the VLBA 20-cm map \citep{gupta2012} to its arcsecond-scale flux density. Assuming 
$\kappa$ = 1 and the \nhi\ estimated from SED fitting, we get \ohi\ $\le1\times10^{-8}$.
\subsection{\hi\ \21\ absorption towards J1501$+$1822}
\label{sec_j1501}
\begin{figure*}
\subfloat[J1501$+$1822]{\includegraphics[width=0.35\textwidth, angle=90]{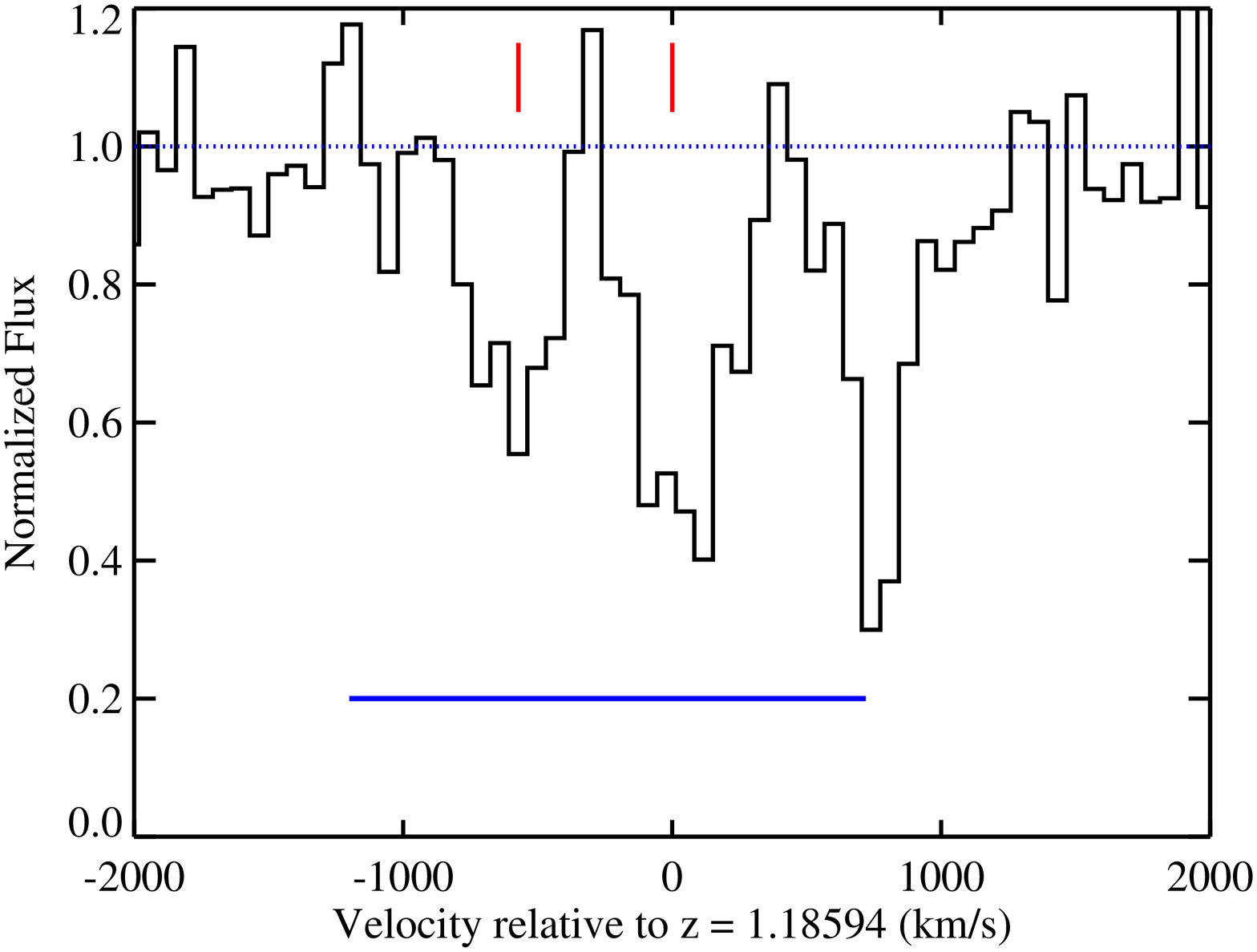} } 
\subfloat[J1521$+$5508]{\includegraphics[width=0.35\textwidth, angle=90]{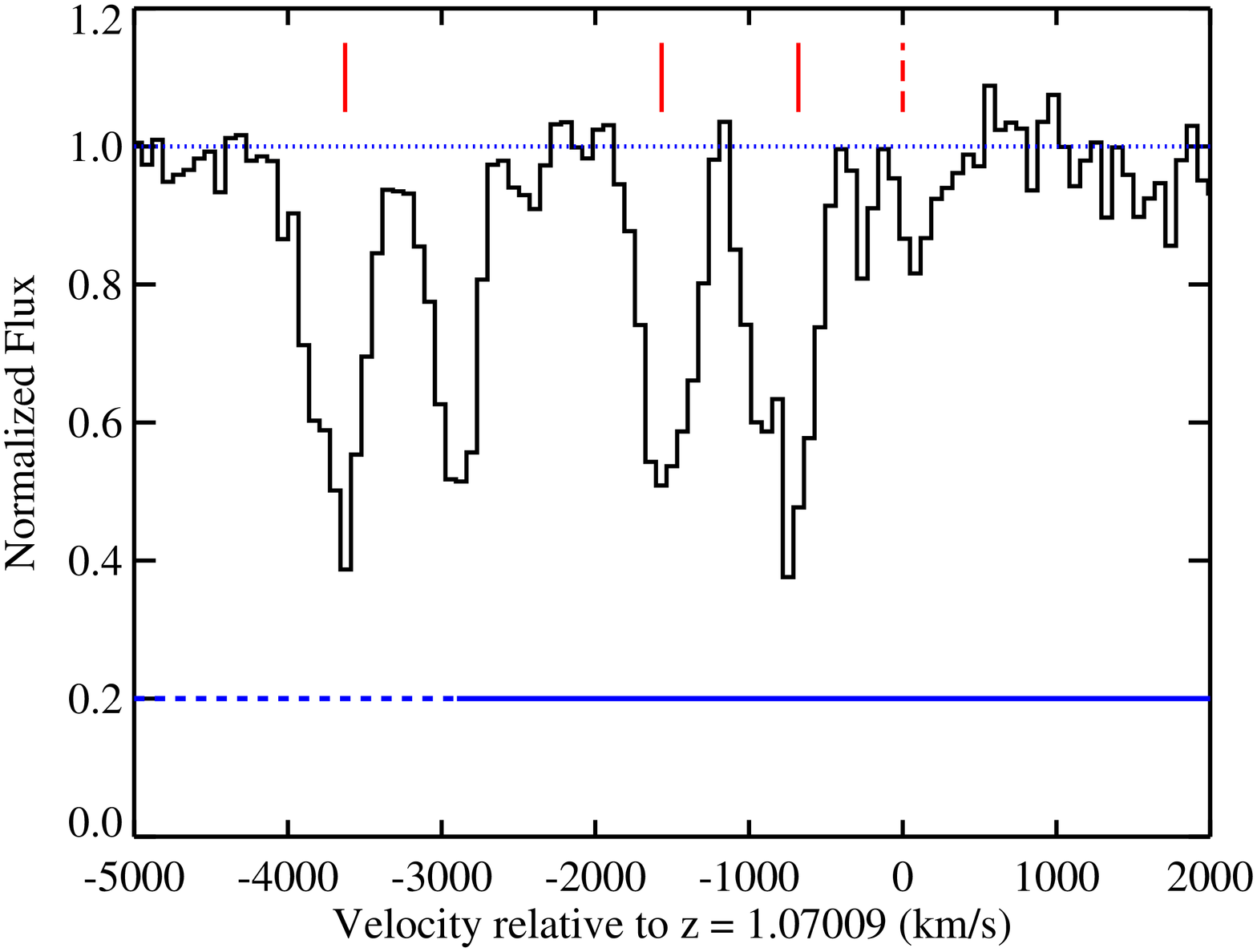} } 
\caption{Normalized SDSS spectra of (a) J1501$+$1822 and (b) J1521$+$5508, showing the associated \mgii\ absorption at $z\sim1.1$.
The velocity scale is with respect to the \zem\ of the quasars. The solid vertical lines mark the position of the different \mgii\,\l2796 
absorption lines. The dashed vertical line in panel (b) marks the redshift of the \hi\ \21\ absorption, which also coincides with the systemic 
redshift of the quasar. The solid horizontal lines demarcate the spectral coverage over which we have clean \hi\ \21\ absorption spectra.
The dashed horizontal line demarcates spectral range affected by RFI.
}
\label{fig:mgii}
\end{figure*}

The SDSS spectrum of the quasar J1501$+$1822 is intriguing as it is devoid of any prominent broad emission lines. We identify the redshift
of the quasar from \oii\ emission line and associated \mgii\ absorption line as \zem\ = 1.1859 $\pm$ 0.0007. There is a second \mgii\ absorption
$\sim$570\,\kms\ blueward at $z$=1.18177 (see panel (a) of Fig.~\ref{fig:mgii}). The \mgii\,\l2796 transition at $z$=1.18594 is blended with 
the \mgii\,\l2803 transition at $z$=1.18177, leading to \wmg\ $\le2.3$\,\AA. For the $z$=1.18177 absorption, \wmg\ $= 1.23 \pm 0.20$\,\AA. 
Weak \feii\ absorption lines are also detected from these two systems, with \wfe\ $= 0.28 \pm 0.15$\,\AA\ at $z=1.18177$ and \wfe\ $= 0.65 \pm 0.14$\,\AA\ 
at $z$=1.18594. For both the systems, we get a $3\sigma$ upper limit on REW of \mgi\ $\le0.27$\,\AA. From integrating the \oii\ emission 
line, we get \oii\ luminosity of $4.0 \pm 0.3 \times10^{42}$\,erg~s$^{-1}$ (uncorrected for dust). Quasars with associated \mgii\ absorption 
have been found to show significant excess in \oii\ emission, which could be due to either enhanced star formation in the quasar host or the 
AGN itself \citep{shen2012,khare2014}. If the \oii\ emission in this case is caused by starburst, then the star formation rate (SFR), uncorrected
for dust, would be $27\pm2$\,M$_\odot$~yr$^{-1}$ \citep[following][]{kewley2004}. 

\begin{figure*}
\subfloat[J1501$+$1822]{\includegraphics[width=0.35\textwidth, angle=90]{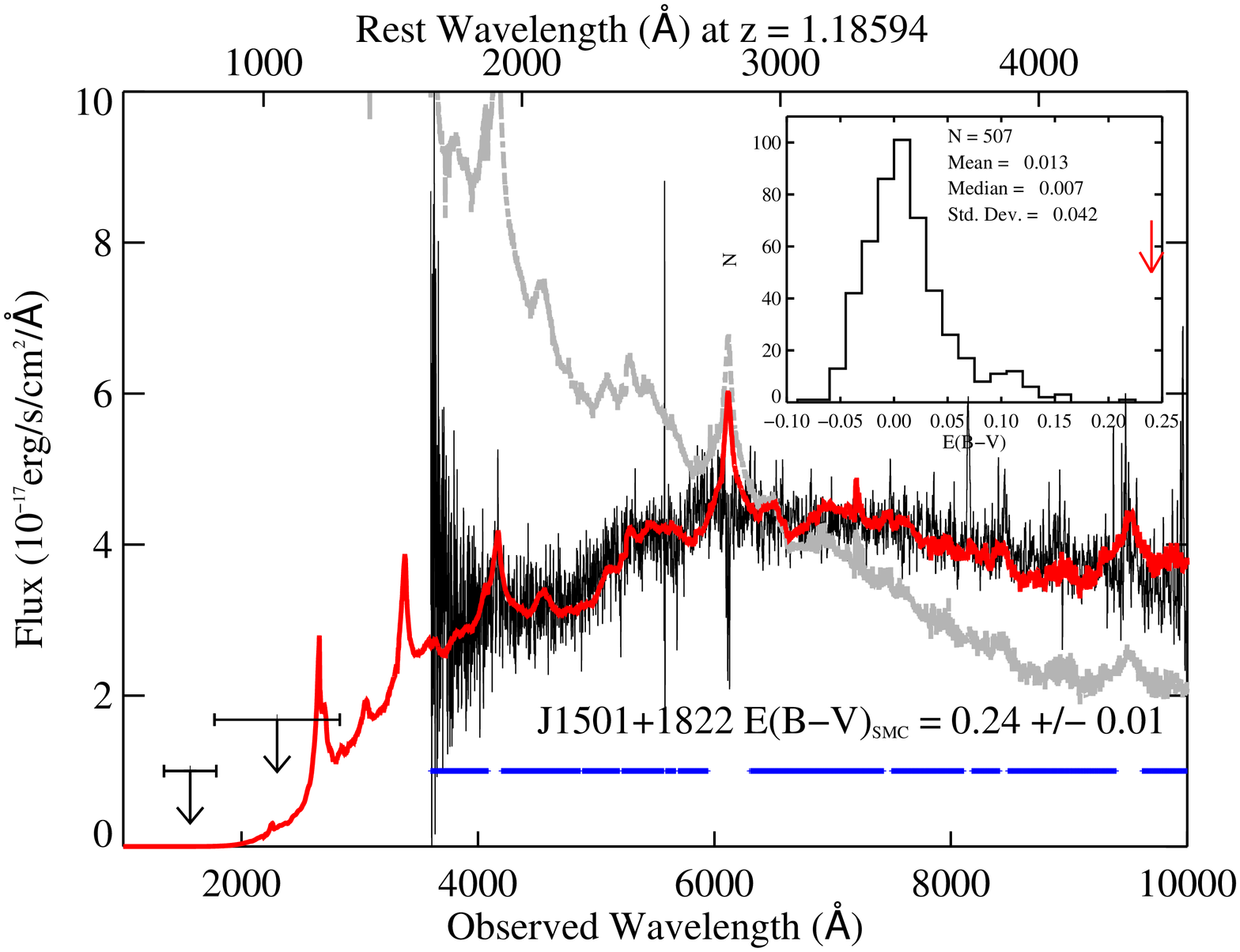} } 
\subfloat[J1521$+$5508]{\includegraphics[width=0.35\textwidth, angle=90]{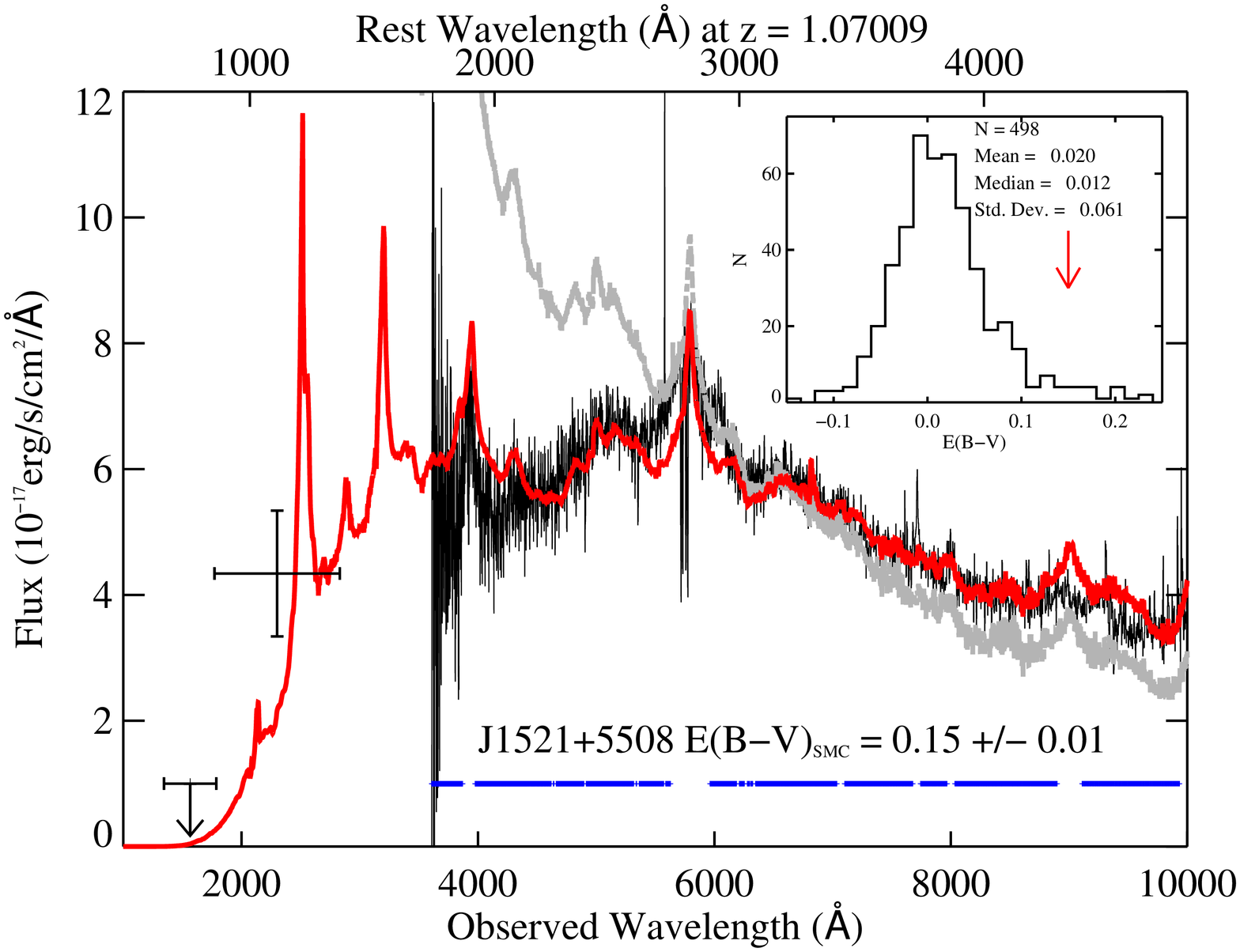} } 
\caption{SED fit to the spectra of (a) J1501$+$1822 and (b) J1521$+$5508. The best-fitting SED (red), obtained from the SMC extinction curve, 
is overplotted on the SDSS spectrum (black) of the quasar in each panel. The non-reddened quasar composite spectrum \citep{selsing2016,telfer2002} 
is shown in grey dashed lines. The points with error bars indicate the ultraviolet flux estimated from GALEX photometry. The horizontal lines indicate 
the spectral regions included in the fitting process. The best-fitting \ebv\ is indicated in each panel. The inset in each panel shows the distribution 
of \ebv\ for a control sample of SDSS quasars (see Section~\ref{sec_j1501} for details), along with the number of quasars, mean, median and standard 
deviation of \ebv\ in the control sample. The arrows indicate the \ebv\ of the corresponding quasar.
}
\label{fig:sed}
\end{figure*}

The SDSS spectrum of the quasar is best fit with the SMC Bar extinction law and \ebv\ $= 0.24 \pm 0.01$, assuming that the reddening is due to 
dust associated with the strong \mgii\ absorption at the systemic redshift (see Fig.~\ref{fig:sed}). The error in \ebv\ include the uncertainties 
in the extinction law parameters. Note that there are no strong intervening metal line absorption at $z\ge$0.4 detected in the SDSS spectrum, so
the reddening of the quasar is most likely due to the associated \mgii\ system. We further applied the same SED fitting procedure with SMC Bar 
extinction law to a control sample of non-BAL quasars in SDSS, within $\Delta z = \pm0.1$ of \zem\ and $\Delta r_{mag} = \pm0.5$ of $r_{mag}$ (21) 
of the quasar and having spectra with signal-to-noise ratio $\ge$10. The details of the control sample are given in Fig.~\ref{fig:sed}. The standard 
deviation of the \ebv\ distribution of the control sample reflects the typical systematic error in the SED fitting procedure due to the dispersion 
of the non-reddened quasar SED. In this case, we find that the quasar is reddened at a significance level of $\sim6\sigma$. The best-fitting SED is 
also found to be consistent with the Galaxy Evolution Explorer (GALEX) UV photometry (Fig.~\ref{fig:sed}). Here for the UV part of the wavelength, 
we have taken the composite spectrum of \citet{telfer2002}. From the near-UV (NUV) flux, we obtain the luminosity as $3\times10^{21}$~W~Hz$^{-1}$.

\hi\ \21\ absorption is not detected from this quasar at the redshift of either \mgii\ absorption (see panel (c) of Fig.~\ref{fig:absspectra}). From the 
observed optical depth limit, the $3\sigma$ upper limit on \nhi\ is $3\times10^{19}$ (\ts/100~K) (1/\fchi) \cms\ per 100\,\kms\ linewidth. On the other 
hand, from SED fitting we expect \nhi\ $\sim9\times10^{21}\kappa^{-1}$\,\cms. Here $\kappa$ is the dust-to-gas ratio relative to the SMC. Comparing the 
two \nhi\ estimates, we can constrain ($\kappa$ \ts)$/$\fchi\ $\ge3\times10^4$~K. The radio source is classified as a flat-spectrum radio source \citep{healey2007}, 
with flux of 55~mJy at 8.4~GHz and 97~mJy at 4.8~GHz, and a spectral index of $-0.46$ between 650~MHz and 4.8~GHz. Considering the flat spectral index, 
the radio emission is likely to be compact, but in the absence of high resolution images the covering factor is undetermined. 
\subsection{\hi\ \21\ absorption towards J1521$+$5508}
\label{sec_j1521}
The red quasar J1521$+$5508 shows broad emission lines in its SDSS spectrum at \zem\ = 1.0701 $\pm$ 0.0004. This is an interesting quasar which has three 
blueshifted \mgii\ absorption at $z=$ 1.04504,  1.05926 and 1.06540 (see panel (b) of Fig.~\ref{fig:mgii}). The \mgii\ absorption lines could be arising 
from strong outflows from the quasar or from an over-density of galaxies clustering around the quasar. The \mgii\,\l2796 transition at $z$=1.06540 is 
blended with the \mgii\,\l2803 transition at $z$=1.05926. The \wmg\ for the three absorption systems at $z$= 1.04504,  1.05926 and 1.06540 are $2.15 \pm 0.12$\,\AA, 
$1.5 \pm 0.1$\,\AA\ and $\le2.3$\,\AA, respectively. The corresponding \wfe\ for the three systems are $\le0.37$\,\AA, $0.70 \pm 0.11$\,\AA\ and $\le0.37$\,\AA,
respectively. The $3\sigma$ upper limit on REW of \mgi\ is $\le0.29$\,\AA\ for all three systems. The \oii\ emission line detected in the SDSS spectrum 
at $z$=1.07009 gives a dust-uncorrected luminosity of $1.6 \pm 0.3 \times10^{42}$\,erg~s$^{-1}$, and SFR of $11\pm2$\,M$_\odot$~yr$^{-1}$ \citep[following][]{kewley2004}, 
if the emission is due to starburst in the host galaxy. The UV luminosity of this quasar, based on the GALEX NUV photometry, is $7\times10^{21}$~W~Hz$^{-1}$.
This quasar is also identified as a candidate $\gamma$-ray blazar of mixed class based on its WISE mid-infrared colours \citep{dabrusco2014}.

Performing SED fitting of the quasar SDSS spectrum, we find that both SMC Bar and SMC Wing extinction laws give best fit to the quasar, with \ebv\ 
$= 0.15 \pm 0.01$ and $0.20 \pm 0.01$, respectively. However, the fit obtained from SMC Bar extinction law is more consistent with the GALEX UV 
photometry (see Fig.~\ref{fig:sed}). Hence, we adopt this fit for the discussion here. Based on SED fitting to a control sample of SDSS quasars 
(similar to Section~\ref{sec_j1501}), the reddening of this quasar is significant at $2.5\sigma$ level. The \nhi\ derived from the SED fitting is
$\sim5\times10^{21}\kappa^{-1}$\,\cms. Note that the reddening is assumed to be due to dust present in the quasar host galaxy. Some of the reddening 
could also be due to dust associated with the blueshifted \mgii\ absorption systems, in which case the above estimated \ebv\ and \nhi\ values would be 
upper limits. SED fits that assume reddening due to dust in the quasar host galaxy or in the \mgii\ systems have similar \chin\ values. Note that there 
are no intervening \mgii\ systems detected over 0.4$\le z\le$1.0 in the SDSS spectrum.

Initially we observed this quasar with a bandwidth of 16 MHz and a resolution of $\sim14$\,\kms, which covered the \hi\ \21\ line at the redshift of the three 
\mgii\ systems and at the systemic redshift of the quasar. We did not detect any absorption at $z=$ 1.05926 and 1.06540, and the frequency ranges around $z=$
1.04504 were affected by poor bandpass stability and RFI. From this spectrum, we put a $3\sigma$ upper limit on \nhi\ $\le5\times10^{19}$ (\ts/100~K) (1/\fchi)
\cms\ per 100\,\kms\ near $z$= 1.05926 and 1.06540. In addition, we detected a tentative absorption at the quasar systemic redshift of 1.07009. 

We carried out follow-up observation of this system, with a 2 MHz band centred at the tentative detection and a resolution of $\sim$2\,\kms. The resultant spectrum 
confirmed the detection, with the peak optical depth detected at $\sim4\sigma$ and the integrated optical depth at $\sim7\sigma$. We show the absorption spectrum 
rebinned to 10\,\kms\ resolution in panel (d) of Fig.~\ref{fig:absspectra}. The \hi\ absorption coincides with the systemic redshift of the quasar and is redshifted 
by $\sim$3628\,\kms, $\sim$1568\,\kms\ and $\sim$679\,\kms\ from the three \mgii\ absorption systems. Note that we do not detect any strong \mgii\ absorption
(\wmg\ $\le0.6$\,\AA) at the redshift of the \hi\ absorption. The total optical depth of the absorption translates to \nhi\ $= (1.2 \pm 0.2) \times 10^{20}$ 
(\ts/100~K) (1/\fchi) \cms. We fitted a single component Gaussian to the absorption profile with a FWHM of 60\,\kms. The velocity width which contains 90\% of 
the optical depth is 75\,\kms.

Combining the observed optical depth and the \nhi\ derived from SED fitting, gives us ($\kappa$ \ts)$/$\fchi\ $\sim4500$~K. Based on the 4.85 GHz flux of 45 mJy
\citep{gregory1991} and the observed flux in our uGMRT data, we estimate a spectral index of $-0.75$ for the radio source. Since no sub-arcsecond-scale image exists
for this radio source, \fchi\ cannot be constrained. The dust-to-gas ratio in this system could be higher than that observed in the SMC, i.e. $\kappa$ could be
greater than unity, as is possible in high-$z$ absorbers due to different grain chemistry or size \citep[see][]{shaw2016,noterdaeme2017,dutta2017b,rawlins2018}. 
This is similar to what is inferred from the lack of \hi\ \21\ absorption towards the red quasar J0956$+$4046 \citep[see Section~\ref{sec_sample1} and][]{gupta2012}.
The above estimate on ($\kappa$ \ts)$/$\fchi\ can thus be taken as an upper limit on \ts.

%
\section{Discussion} 
\label{sec_discussion}
We have started searching for \hi\ \21\ and OH \18\ absorption lines at $z\ge1$ towards red quasars in the newly available frequency ranges offered by the uGMRT.
We have presented here the results for our search for OH \18\ absorption from two intervening strong \mgii\ systems at $z=1.3$ and \hi\ \21\ absorption from 
two associated strong \mgii\ systems at $z=1.1$, all towards quasars showing reddening in their optical spectra [\ebv\ $\sim$0.1-0.4].
\subsection{OH \18\ absorption towards red quasars}
\label{sec_discussion_oh}
The quasars J0850$+$5159 and J0852$+$3435, although weak in their radio emission ($\sim$60\,mJy at 1.4~GHz), were promising candidates for follow-up 
OH \18\ absorption observations because of the presence of strong metal absorption (\wmg\ $\gtrsim$3\AA), 2175\,\AA\ extinction bump, and strong dust 
reddening [\ebv\ $\sim$0.3-0.4]. The \nhi\ values [$\sim$(5-7)$\times10^{21}$\,\cms] derived from SED fitting of the SDSS spectra are also high \citep{srianand2008}, 
though there is no evidence for strong dependence of the OH detection rate on \nhi\ among the known Galactic and extragalactic OH absorbers \citep{gupta2018}. 
The $3\sigma$ upper limits achieved on \noh\ [$\sim$(1-2)$\times10^{14}$\,\cms] should be sensitive to detect the higher end of the \noh\ detected in our Galaxy 
\citep{li2018,rugel2018}. The constraints on the \ohi\ abundance ratio [$\le$(1-4)$\times10^{-8}$] are lower than the median \ohi\ $\sim10^{-7}$ observed in our 
Galaxy \citep{li2018,rugel2018}. This could indicate towards an evolution in the physical conditions and abundance of molecular gas at $z>1$. However, we note 
that the scatter in the Galactic measurements is large, and that dust-to-gas ratios greater than that in the LMC would push the above derived limits higher. 
In both these systems, we can most likely rule out the presence of dense molecular gas. The presence of dust in these systems means that we cannot rule out 
the presence of molecular gas completely along these sightlines. We are hence probably still not reaching the sensitivity required to detect the diffuse 
molecular gas, i.e. with \noh\ $\le10^{14}$\,\cms. The upcoming MeerKAT Absorption Line Survey \citep[MALS;][]{gupta2016} will specifically target bright 
high-$z$ quasars to reach OH column density sensitivities below this ($\sim5\times10^{13}$\,\cms) over $0<z<2$ \citep{gupta2018}. 
\subsection{\hi\ \21\ absorption towards red quasars}
\label{sec_discussion_hi}
We have detected \hi\ \21\ absorption at the systemic redshift of the quasar J1521$+$5508, with \nhi\ $=(1.2\pm0.2)\times10^{20}$ (\ts/100~K) (1/\fchi) \cms.
However, we do not detect \hi\ absorption from the blueshifted strong \mgii\ absorption systems along this line-of-sight [\nhi\ $\le5\times10^{19}$ (\ts/100~K) (1/\fchi) \cms].
We also do not detect \hi\ absorption from the strong \mgii\ systems associated with the quasar J1501$+$1822 [\nhi\ $\le3\times10^{19}$ (\ts/100~K) (1/\fchi) \cms].
The reasons for non-detection of associated \hi\ \21\ absorption in general are the absence of neutral gas due to ionization by high luminosity-AGN, 
higher spin temperature in the vicinity of AGN, and extended background radio structure and small-scale structure in the absorbing gas itself leading 
to small covering factor. The two quasars discussed here show the presence of strong \mgii\ absorption and dust reddening, which indicate the presence 
of neutral gas in the vicinity of the quasars. The \nhi\ estimates inferred from SED fitting, assuming that the reddening is due to dust associated with 
the quasars, are $\sim$(5-9)$\times10^{21}$\,\cms. If part of the reddening is caused due to dust present in the blueshifted \mgii\ absorption systems, the 
\nhi\ estimates would be upper limits. Additionally, based on the \wmg\ and \wfe, the probability of \nhi\ being greater than $10^{19}$ and $10^{20}$ \cms, 
is 92\% and 51\%, respectively \citep[see][for details]{dutta2017b}. Further, the UV luminosity of these quasars is below the critical luminosity believed 
to ionize neutral gas completely \citep{curran2008,curran2012}. Hence, we can rule out the absence of neutral gas as the reason behind the lack of 
\hi\ \21\ absorption.

The relatively flat radio spectrum of these two quasars implies compact structure, though in absence of sub-arcsecond-scale images, the 
background radio structure and exact gas covering factor are uncertain. We are probing absorption over scales of $\sim30-40$~kpc at the 
redshift of the quasars\footnote{Adopting a flat $\Lambda$-cold dark matter cosmology with $H_{\rm 0}$ = 70\,\kms~Mpc$^{-1}$ and $\Omega_M$ 
= 0.30.}. It is likely that the neutral gas is present in small clumps along these sightlines, and we are not detecting \hi\ \21\ absorption 
from all the clumps due to difference in the optical and radio sightlines. This would explain the detection of \hi\ \21\ absorption at the 
systemic redshift of the quasar J1521$+$5508 and non-detection at the redshift of the blueshifted \mgii\ absorption. Indeed, the absence of 
strong \mgii\ absorption at the redshift of the \hi\ \21\ absorption implies that the radio sightline is probing a different volume of gas 
than the optical sightline. Differences in \hi\ \21\ and metal absorption profiles, with the \hi\ \21\ absorption not always coinciding with 
the strongest metal component, have been observed even in intervening systems with high resolution optical spectra available \citep[e.g.][]{gupta2009,rahmani2012,dutta2015}. 
While no strong correlation has been found between line width of intervening \hi\ \21\ absorption and linear size of radio sources from 
subarcsecond-scale images, the largest velocity widths are seen towards radio sources with extended structure at arcsecond-scales \citep{gupta2012}.
This can be explained if the absorbing gas is patchy, which is consistent with the above scenario that gives rise to differences in \hi\ and 
metal absorption.

Next, we discuss the detection rate of \hi\ \21\ absorption in red quasars and in particular the detection rate of associated absorption at $z\ge$1.
As outlined in Section~\ref{sec_sample2}, the two quasars, J1501$+$1822 and J1521$+$5508, were selected on the basis of their SDSS red colours and 
strong \mgii\ absorption. This selection procedure resulted in a total sample of eight red quasars (with associated and intervening absorbers), out 
of which six had been previously observed, with two detections. Hence, including the two quasars presented here, the detection rate of \hi\ \21\ 
absorption in this sample of red quasars is 38$^{+36}_{-20}$\%\footnote{Poissonian errors computed using tables of \citet{gehrels1986}.}. 

Considering only associated absorption, two out of three red quasars show \hi\ \21\ absorption $-$ towards J1521$+$5508 reported here, and towards
3C\,190 ($z$=1.2) \citep{ishwarachandra2003}. It is interesting to note that, similar to the velocity offset in \hi\ and \mgii\ absorption towards 
J1521$+$5508, the peak \hi\ optical depth in case of the absorption towards 3C\,190 also does not coincide in velocity with the peak of the \mgii\ 
absorption. Even though the uncertainties are large due to the small number of red quasars searched to date at $z\ge$1, the detection rate of 
associated \hi\ \21\ absorption in red quasars is higher than the typical detection rate ($\le$10\%) in samples of $z\ge$1 AGNs selected without 
any condition on reddening or the presence of metal absorption. For example, from the compilation presented in \citet{aditya2018a} and \citet{aditya2018b}, 
associated \hi\ \21\ absorption has been detected in 4 out of 56 flat-spectrum and gigahertz-peaked-spectrum radio sources at $z\ge$1. SDSS spectrum 
is available for one of the four sources with detection (TXS\,1200$+$045) and we do not detect any strong associated \mgii\ absorption (\wmg\ $\le$1\,\AA) 
within $\sim$3000\,\kms\ of the systemic redshift of the quasar. The optical spectra available for two other sources (TXS\,1245$-$197 and TXS\,1954$+$513) 
in the literature \citep{alighieri1994,lawrence1996} do not show signature of associated \mgii\ absorption. However, these are very low-resolution 
($\sim$10\,\AA) spectra from which we cannot place any strong constraints. Note that $\sim$60\% of the \hi\ absorption spectra reported in \citet{aditya2018a} 
and \citet{aditya2018b} are not sensitive enough to detect the absorption towards J1521$+$5508 reported here. For 17 of the sources with 
\hi\ \21\ non-detection, SDSS spectra are available, and we do not detect \mgii\ absorption (\wmg\ $\le$1\,\AA) within $\sim$3000\,\kms\ of the 
systemic redshift. Similarly, for the \hi\ \21\ non-detections in $z\sim1-2$ compact radio sources reported in the recent study of \citet{grasha2019}, 
no associated \mgii\ absorption is detected in the available SDSS spectra. Hence, the lack of \hi\ \21\ absorption in these cases is consistent with 
the absence of metal-rich neutral gas in the vicinity of the AGNs.

Next, we consider all quasars with associated \mgii\ absorption at $z\ge1$ in SDSS spectra that have been searched for \hi\ \21\ absorption. 
In addition to the three systems in our red quasar sample, there are five quasars that do not fulfill the optical colour or radio flux criteria
of our sample (Section~\ref{sec_sample2}), but show \mgii\ absorption at $\le$3000\,\kms\ of the systemic redshift. Three out of these five have 
been detected in \hi\ \21\ absorption $-$ J010826.84$-$003724.1 \citep[$z$=1.375;][]{gupta2009}, J114521.32$+$045526.7 \citep[$z$=1.345;][]{kanekar2009},
J091927.61$+$014603.0 \citep[$z$=1.286;][]{dutta2017b}. The peak \hi\ optical depth of the absorption is again offset (up to $\sim$100\,\kms) 
from the peak \mgii\ absorption in these systems. Two of these quasars have VLBA images available, which show structures in the radio emission at 
subarcsecond-scales, with \fchi\ $\sim$0.3-0.5 \citep{gupta2012}. This supports our above argument that the neutral gas near these AGNs have 
structures at small scales that give rise to different absorbing components in the optical and radio sightlines. The total detection rate of 
associated \hi\ \21\ absorption in the above \mgii-selected systems is 62$^{+0.38}_{-0.27}$\% (57$^{+0.43}_{-0.27}$\% considering
only \wmg\ $\ge$1\,\AA\ systems). Thus, the presence of strong associated \mgii\ absorption increases the detection rate of associated \hi\ \21\ 
absorption in $z\ge$1 quasars. Due to small number statistics, we are not able to determine yet whether the presence of reddening signature 
further enhances the detection rate.

\begin{figure}
\includegraphics[width=0.38\textwidth, angle=90]{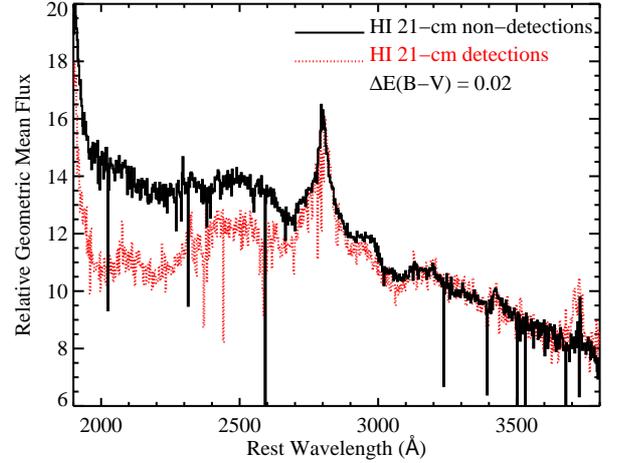} 
\caption{Stacked SDSS spectra (geometric mean) of the $z\ge$1 quasars at rest-frame, with and without detection of associated \hi\ \21\ absorption in red dotted and black solid lines, respectively.
}
\label{fig:stack}
\end{figure}

Similar to what has been found for intervening \mgii\ absorbers \citep{gupta2012,dutta2017b}, we do not find any correlation of the \hi\ \21\ optical 
depth and velocity width with \wmg\ of the associated \mgii\ absorption at $z\ge1$, albeit the number of such absorbers is small (5). However, considering 
the effect of reddening alone, we find that the distributions of the colour excess, $\Delta (r-i)$, of the quasars with and without associated \hi\ \21\ 
absorption at $z\ge$1 are different. A two-sided Kolmogorov-Smirnov test gives maximum deviation between the two cumulative distributions as $D_{\rm KS}$ 
= 0.54, with the probability of finding this difference by chance being $P_{\rm KS}$ = 0.03. The effect of reddening due to dust can be seen more clearly 
from the geometric mean stacked SDSS spectra, at the rest-frame of the quasar, of the associated \hi\ \21\ absorbers (6) and non-absorbers (26) at 
$z\ge$1\footnote{For this analysis we consider all the quasars searched for associated \hi\ \21\ absorption at $z\ge$1 that have SDSS spectra available.} 
(Fig.~\ref{fig:stack}). The quasars which show associated \hi\ \21\ absorption tend to be more reddened on average, with the stacked spectrum of quasars 
showing associated \hi\ \21\ absorption having a differential reddening of $\Delta$\ebv\ = 0.02 with respect to the stacked spectrum of quasars not detected 
in associated \hi\ \21\ absorption. Therefore, there are indications of both strong \mgii\ absorption and optical reddening leading to higher occurrence of 
associated \hi\ \21\ absorption in $z\ge$1 quasars. Further searches towards reddened quasars along with characterization of the parsec-scale radio structure 
through VBLA imaging are the way forward towards understanding the incidence, distribution and physical conditions of neutral gas in high-$z$ quasars. \\
%
%
%
%
%

\noindent \textbf{ACKNOWLEDGEMENTS} \newline \newline
\noindent 
We thank the anonymous reviewer for their helpful comments and suggestions.
RD acknowledges support from the Alexander von Humboldt Foundation.
We thank the staff at GMRT for their help during the observations. 
GMRT is run by the National Centre for Radio Astrophysics of the Tata Institute of Fundamental Research.
Funding for SDSS-III has been provided by the Alfred P. Sloan Foundation, the Participating Institutions, the National Science Foundation, 
and the U.S. Department of Energy Office of Science. The SDSS-III web site is http://www.sdss3.org/.
SDSS-III is managed by the Astrophysical Research Consortium for the Participating Institutions of the SDSS-III Collaboration 
including the University of Arizona, the Brazilian Participation Group, Brookhaven National Laboratory, Carnegie Mellon University, 
University of Florida, the French Participation Group, the German Participation Group, Harvard University, the Instituto de Astrofisica 
de Canarias, the Michigan State/Notre Dame/JINA Participation Group, Johns Hopkins University, Lawrence Berkeley National Laboratory, 
Max Planck Institute for Astrophysics, Max Planck Institute for Extraterrestrial Physics, New Mexico State University, New York University, 
Ohio State University, Pennsylvania State University, University of Portsmouth, Princeton University, the Spanish Participation Group, 
University of Tokyo, University of Utah, Vanderbilt University, University of Virginia, University of Washington, and Yale University. 
%
%
\def\aj{AJ}%
\def\actaa{Acta Astron.}%
\def\araa{ARA\&A}%
\def\apj{ApJ}%
\def\apjl{ApJ}%
\def\apjs{ApJS}%
\def\ao{Appl.~Opt.}%
\def\apss{Ap\&SS}%
\def\aap{A\&A}%
\def\aapr{A\&A~Rev.}%
\def\aaps{A\&AS}%
\def\azh{A$Z$h}%
\def\baas{BAAS}%
\def\bac{Bull. astr. Inst. Czechosl.}%
\def\caa{Chinese Astron. Astrophys.}%
\def\cjaa{Chinese J. Astron. Astrophys.}%
\def\icarus{Icarus}%
\def\jcap{J. Cosmology Astropart. Phys.}%
\def\jrasc{JRASC}%
\def\mnras{MNRAS}%
\def\memras{MmRAS}%
\def\na{New A}%
\def\nar{New A Rev.}%
\def\pasa{PASA}%
\def\pra{Phys.~Rev.~A}%
\def\prb{Phys.~Rev.~B}%
\def\prc{Phys.~Rev.~C}%
\def\prd{Phys.~Rev.~D}%
\def\pre{Phys.~Rev.~E}%
\def\prl{Phys.~Rev.~Lett.}%
\def\pasp{PASP}%
\def\pasj{PASJ}%
\def\qjras{QJRAS}%
\def\rmxaa{Rev. Mexicana Astron. Astrofis.}%
\def\skytel{S\&T}%
\def\solphys{Sol.~Phys.}%
\def\sovast{Soviet~Ast.}%
\def\ssr{Space~Sci.~Rev.}%
\def\zap{$Z$Ap}%
\def\nat{Nature}%
\def\iaucirc{IAU~Circ.}%
\def\aplett{Astrophys.~Lett.}%
\def\apspr{Astrophys.~Space~Phys.~Res.}%
\def\bain{Bull.~Astron.~Inst.~Netherlands}%
\def\fcp{Fund.~Cosmic~Phys.}%
\def\gca{Geochim.~Cosmochim.~Acta}%
\def\grl{Geophys.~Res.~Lett.}%
\def\jcp{J.~Chem.~Phys.}%
\def\jgr{J.~Geophys.~Res.}%
\def\jqsrt{J.~Quant.~Spec.~Radiat.~Transf.}%
\def\memsai{Mem.~Soc.~Astron.~Italiana}%
\def\nphysa{Nucl.~Phys.~A}%
\def\physrep{Phys.~Rep.}%
\def\physscr{Phys.~Scr}%
\def\planss{Planet.~Space~Sci.}%
\def\procspie{Proc.~SPIE}%
\let\astap=\aap
\let\apjlett=\apjl
\let\apjsupp=\apjs
\let\applopt=\ao
\bibliographystyle{mnras}
\bibliography{mybib}
\bsp
\label{lastpage}
\end{document}